\def\aap{A\&A}

\def\apj{ApJ}
\def\apjl{ApJL}
\def\apss{ApSS}

\def\mnras{MNRAS}

\def\deg{$^\circ$}
\documentclass[letter]{aa}  
\usepackage{graphicx}
\usepackage{txfonts}

\begin{document}
%

\title{A silicate disk in the heart of the Ant}

\authorrunning{Chesneau et al.}
\titlerunning{A silicate disk in the heart of the Ant}

   \author{O. Chesneau
          \inst{1}
          \and F. Lykou
           \inst{2}
          \and
          B. Balick\inst{3}
          \and
          E. Lagadec
           \inst{2}
          \and
          M. Matsuura
          \inst{4}
          \and
          N. Smith\inst{5}
          \and
          A. Spang\inst{1}
          \and
          S. Wolf\inst{6}
          \and
          A.A. Zijlstra
           \inst{2}
           \fnmsep\thanks{Based on observations made with the Very Large Telescope Interferometer at Paranal Observatory under program 077.D-0193}
          }

   \offprints{O. Chesneau}

   \institute{Observatoire de la C\^{o}te d'Azur-CNRS-UMR 6203, Dept Gemini,
Avenue Copernic, F-06130 Grasse, France\\
              \email{olivier.chesneau@ob-azur.fr}%
              \and
University of Manchester, School of Physics \& Astronomy, P.O. Box 88, Manchester M60 1QD, UK
\and
Astronomy Department, Box 351580, University of Washington, Seattle WA 98195
\and
National Astronomical Observatory of Japan, Osawa 2-21-1, Mitaka, Tokyo 181-8588, Japan
\and
Astronomy Department, University of California, 601 Campbell Hall, Berkeley CA 94720
\and
Max-Planck-Institut f\"{u}r Astronomie,
   K\"{o}nigstuhl 17, D-69117 Heidelberg, Germany
}

   \date{Received ;accepted }

 
  \abstract
   {}
   {We aim at getting high spatial resolution information on the dusty core of bipolar planetary nebulae to directly constrain the shaping process.}
   {We present observations of the dusty core of the extreme bipolar planetary
nebula Menzel\,3 (Mz\,3, Hen 2-154, the Ant) taken with the mid-infrared interferometer
MIDI/VLTI and the adaptive optics NACO/VLT.}
   {The core of Mz\,3 is clearly resolved with MIDI in the interferometric mode, whereas it is unresolved from the Ks to the N bands with single dish 8.2m observations on a scale ranging from 60 to 250mas. A striking dependence of the dust core size with the PA angle of the baselines is observed, that is highly suggestive of an edge-on disk whose major axis is perpendicular to the axis of the bipolar lobes. The MIDI spectrum and the visibilities of Mz\,3 exhibit a clear signature of amorphous silicate, in contrast to the signatures of crystalline silicates detected in binary post-AGB systems, suggesting that the disk might be relatively young. We used radiative-transfer
Monte Carlo simulations of a passive disk to constrain its geometrical and physical parameters. Its inclination (74\deg$\pm$3\deg) and position angle (5\deg$\pm$5\deg) are in accordance with the values derived from the study of the lobes. The inner radius is 9$\pm$1AU and the disk is relatively flat. The dust mass stored in the disk, estimated as $1 \times 10^{-5}M_{\odot}$, represents only a small fraction of the dust mass found in the lobes and might be a kind of relic of an essentially polar ejection process. }
   {}

                
   \maketitle
%

\section{Introduction}
It is commonly accepted that disks can be an essential ingredient to the shaping of planetary nebulae, but the spatial resolution of astronomical instruments is usually not sufficient for detecting and studying these disks. The geometry of the disk and the mass stored are key parameters for constraining the models of nebula formation, and for tracing back the evolution of the central star (\cite{2002ARA&A..40..439B}). Its inner edge, as seen from the star, could be thick and dense enough to collimate stellar winds into lobes, and knowing its geometry enables us to better understand the distribution of illuminated and shadowed regions in the extended nebula. 
The VLT Interferometer is a powerful `disk hunter' with its two instruments: AMBER operating in the near-IR (\cite{2007A&A...464....1P}) and MIDI in the mid-IR (\cite{2003Ap&SS.286...73L}) providing a typical spatial resolution of 2 and 10mas, respectively. It must be stressed that the number of visibility measurements usually recorded do not allow any image reconstruction from the interferometric observations because this would be too time-consuming. Therefore, the interpretation uses models that depend critically on the information already gathered on the sources by other techniques. As an example, a disk was detected by HST and MIDI in the center of the complex multipolar nebula CPD-56$^\circ$8032 (\cite{2002ApJ...574L..83D}, \cite{2006A&A...455.1009C}). The disk exhibits a bright and extended (on an interferometric scale, $\sim$70mas) inner rim whose PA and inclination do not seem directly related to extended structures in the HST images. Definitely, the disk detection based on three visibility measurements provided a lot of  information, but also demonstrated the need for an interferometric imaging campaign involving wide $uv$ coverage for this complex object.

In contrast, the Butterfly (M2-9) and the Ant (Mz\,3, Hen 2-154) nebulae were chosen as perfectly suited targets to a snapshot study with the VLTI, as they are among the most impressive bipolar PNs, highly collimated on large scales with tightly pinched waists as seen in Hubble Space Telescope (HST) images (\cite{2002ARA&A..40..439B} and references therein; \cite{2003MNRAS.342..383S}; \cite{2005AJ....129..969S}; \cite{2005AJ....130..853S}). For these targets, assuming that the putative disk hidden in their center shares the symmetry of the extended nebula constitutes a good starting point for preparing and interpreting the VLTI observations. The nebula Mz\,3 shows complex structures revealing different stages of mass-loss events that share the same axis of symmetry (\cite{2004A&A...426..185S}, \cite{2004AJ....128.1694G}). Of importance is also the fact that M2-9 and Mz\,3 are spectroscopic twins at visual and near-IR wavelengths. Each nucleus shows a rich optical spectrum with the various iron and helium lines, that are usually associated with a dense circumnuclear cocoon which reprocesses most of the stellar light emitted in our line of sight that hides any close companion from direct view. Each object requires an ionization source with a temperature on the order of 30,000\, K, and the dust mass detected in their nebulae suggests intermediate-mass progenitors (\cite{2005AJ....129..969S}). Published distances to each object vary widely, hovering around 1-2 kpc. A detailed imaging study of the dust in the core of these nebulae has been presented recently by Smith (2003) and Smith \& Gehrz (2005). The spatial resolution of the mid-IR images was 1$\arcsec$, allowing them to isolate the central engines from the surrounding nebulae. The observed mid-IR SED of the unresolved central sources is flatter than a single gray-body, implying that a range of dust temperatures and an additional component of hot dust are both required to fit the near-IR continuum spectrum.
  This suggests the existence of a circumstellar disk.

The study of Smith \& Gehrz (2005) motivated us to use the unique MIDI/VLTI capabilities for detecting the disks in the center of Mz\,3 and M2-9.
The present Letter is entirely devoted to the disk detected in the center of Mz\,3, while M2-9 observations are described in another paper. The observations are presented in Section 2. In Section\,3 we derive the physical parameters of the dusty environment by radiative-transfer models and discuss the results in Section\,4.


\section{Observations}
Mz\,3 was observed with MIDI (\cite{2003Ap&SS.286...73L}; \cite{2007A&A...471..173R}) the mid-infrared recombiner of the Very Large Telescope (VLT). The MIDI/VLTI interferometer operates like a classical Michelson interferometer to combine the mid-IR light (N band, 7.5-13.5$\mu$m) from two VLT Unit Telescopes (8.2m, UTs) or two Auxiliary Telescopes (1.8m, ATs). In our case, only the UTs were used for detecting correlated fluxes as low as about 6Jy from the 30Jy cores of Mz\,3 and M2-9. We used a typical MIDI observing sequence, as described in Ratzka et al. (2007). MIDI can make single-dish acquisition images with a field of view of about 3$\arcsec$ with a spatial resolution of about 0.25$\arcsec$ at 8.7$\mu$m, and provides a flux-calibrated spectrum at small spectral resolution (R=30 in this case) and several visibility spectra from the sources (\cite{2005A&A...435.1043C}, \cite{2005A&A...435..563C}), i.e. spectrally dispersed information on the spatial extension of the source. The MIDI observations of Mz\,3 were performed in May and June 2006 in the so-called SCI\_PHOT mode, meaning that the photometry of each telescope is recorded simultaneously to the fringes. The errors, including the internal ones and the ones from the calibrator diameter uncertainty, range from 3\% to 7\%. The accuracy of the absolute flux calibration is about 10-13\%. The log of the observation is given in Table~\ref{table-log}. We used two different MIDI data reduction packages: MIA
developed at the Max-Planck-Institut f\"ur Astronomie and EWS
developed at the Leiden Observatory (MIA+EWS\footnote{Available at http://www.strw.leidenuniv.nl/$\sim$nevec/MIDI/index.html}, ver.1.5.1).

We also observed Mz\,3 with the near-infrared AO instrument NACO using three broad-band filters centered respectively at $2.18\mu$m (Ks), $3.8\mu$m (L$'$), and 4.78$\mu$m (M$'$). The Ks and L$'$-band data
were taken during the night of 18 August 2006 and the M$'$-band
data were taken during the night of 29 May 2006. These observations were complemented with nearby PSF calibrations using HD~155078 (F5V) in L$'$ and HD156971 (F1III) in M$'$. We used the S27 (Ks) and L27 (L$'$,M$'$) camera mode to obtain a field of view of 28$\arcsec$$\times$28$\arcsec$ and the pixel scale is 27.1mas per pixel. Jittering was used to remove the sky, and chopping was also used for the M$'$ observations. Data reduction was performed using ESO pipeline and self-developed IDL routines. 
 
 \begin{table}
\begin{caption}
{Observing log
}\label{table-log}
\end{caption}
\begin{tabular}{llccc}\hline\hline 
OB   & Time  & Base & \multicolumn{2}{c}{Projected baseline}  \\
 & & & Length  & PA   \\
 &&&[metre] & [degrees] \\
\hline
Mz3-1 & 2006-06-11T23 & U2 - U3 & 46.3 & 1.5  \\
Mz3-2  & 2006-05-15T04 & U2 - U3 & 45.4 & 30.5 \\
Mz3-3  & 2006-05-15T08 & U2 - U3 & 31.4 & 73.8  \\
Mz3-4  & 2006-05-14T08 & U3 - U4 & 60.6 & 149.2  \\
Mz3-5  & 2006-06-11T01 & U3 - U4 & 52.0 & 77.2  \\
Mz3-6  & 2006-05-17T06 & U3 - U4 & 62.5 & 122.1  \\
\hline
\end{tabular}
{Calibrators: HD151249 5.42$\pm$0.06\,mas, HD160668 2.22$\pm$0.1\,mas, HD168723 2.87$\pm$0.13\,mas, HD188512 1.98$\pm$0.1\,mas.}
\end{table}

\begin{figure}
 \centering
\includegraphics[width=8.5cm]{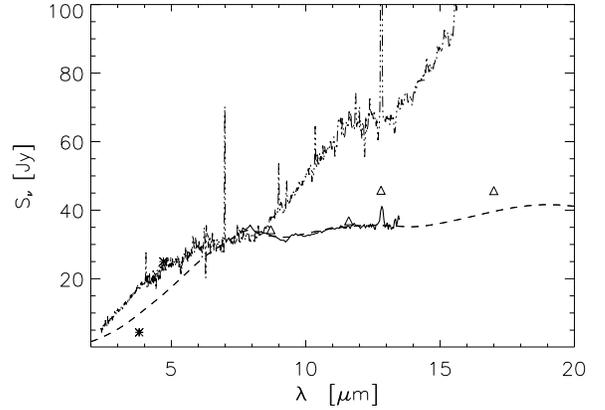}
 \caption[]{MIDI spectrum (thick line) compared with the ISO spectrum (upper curve) and the TIMMI2 photometric measurements of Smith \& Gehrz (2005) with an aperture of 4$\arcsec$ (triangles). The NACO L$'$ and M$'$ bands are also included (stars, aperture 0.4$\arcsec$). The dashed line is the best silicate model with the full aperture (1$\arcsec$). In the N band, the 1$\arcsec$ beam and the MIDI beam flux of the model differ by only a few percent.
\label{fig:siliSED}}
\end{figure}

Figure~\ref{fig:siliSED} shows the ISO and MIDI spectra. The ISO spectra are extensively discussed in Pottasch\,\&\,Surendiranath (2005), and their respective apertures are shown in their Fig.1. We only show the SWS observations with the 14$\arcsec\times$27$\arcsec$ aperture. The flat MIDI spectrum contrasts to the steeply rising ISO flux toward the long wavelengths. Another obvious difference between the two spectra is that the 12.8$\mu$m [NeII] line is barely visible in the MIDI spectrum.

The MIDI flux agrees within the 10\% error bars with the fluxes reported with 4$\arcsec$ apertures by Smith\,\&\,Gehrz (2005, Table 2) at 8.9 and 11.6$\mu$m, but the MIDI flux is significantly lower in the [NeII]. The M$'$ band magnitude of the core agrees within error bars with the ISO flux showing that the whole M$'$ band flux originates in the core (m$_{M'}$=1.9$\pm$0.1). In contrast, the L$'$ band flux of the core is much lower than the ISO curve (m$_{L'}$=4.4$\pm$0.1, aperture 0.4$\arcsec$), due to a strong reddening occurring at this wavelength and the increase in the light diffusion in the lobes. This magnitude value follows the trend in the fluxes measured with decreasing apertures in Cohen et al. (1978).


\section{Physical parameters of the disk}
\begin{figure*}
 \centering
\includegraphics[width=15.cm, height=9.5cm]{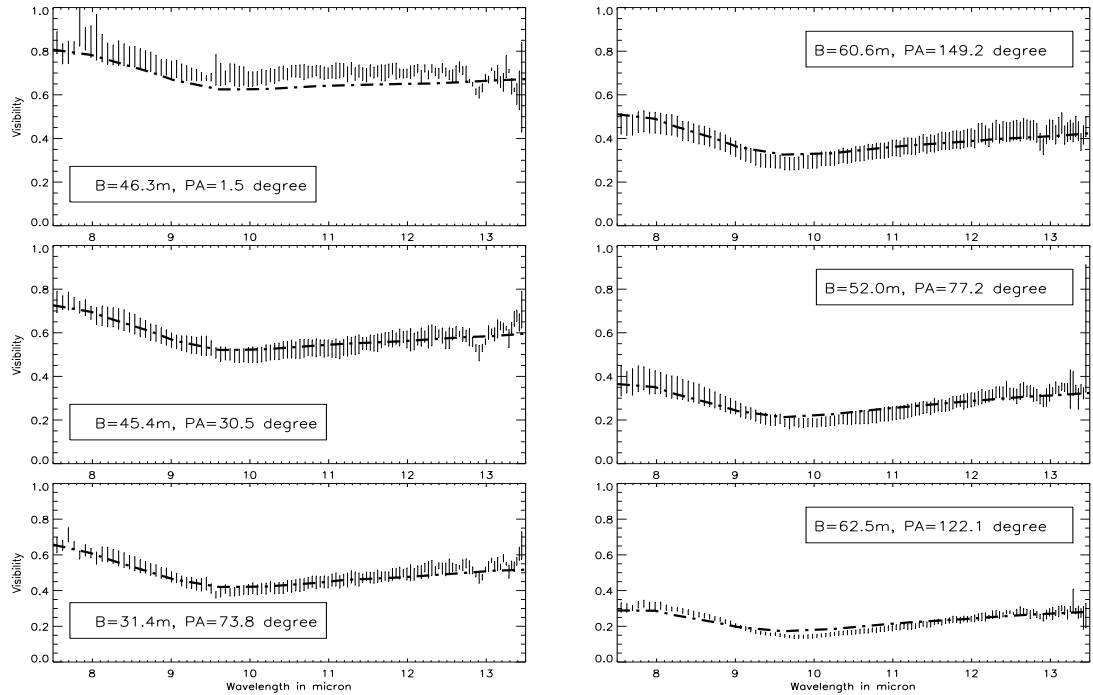}
\hfill
 \caption[]{MIDI visibilities with their error bars, compared with the best silicate model (dashed-dotted lines, see Table\,~\ref{tab:modparam}). The visibility curves are Mz3-1 to Mz3-6 from top to bottom and left to right, with the $\chi^2$ values of the fits of 2.16, 0.30, 0.95, 1.50, 0.88, 4.31. The reduced $\chi^2$ for the full data set is 1.68. The more extended [NeII]12.8$\mu$m line is not taken into account in the model. \label{fig:MIDIvis}}
\end{figure*}

The NACO and MIDI images show that the core is unresolved from the Ks band on a scale of about 60mas to the N band on a scale of about 250mas. 
The large amplitude of variations of the MIDI visibilities versus the baseline position angles can immediately be interpreted in the frame of a very flat structure, whose symmetry axis is roughly aligned with that of the bipolar lobes. In Table\,\ref{table-log} and in Fig.\,\ref{fig:MIDIvis}, the projected baselines are organized from the ones providing the highest visibilities (i.e. smallest extension) at PA close to 0$^\circ$ to the ones with lowest visibilities (i.e. highest extension) distributed around PA=90$^\circ$. A rough estimate of the flux distribution using Gaussian ellipses provides sizes of 10mas x 20mas at 8$\mu$m, and 20mas x 40mas at 13$\mu$m. Another particularly interesting aspect of these visibility curves is their amorphous silicate signature, which is less obvious in the MIDI spectrum and hardly seen in the ISO spectrum. A higher opacity at a particular wavelength will cause a larger radius
to be observed: the interferometer will detect this as a reduced
visibility as compared to the continuum emission. This well-known opacity effect is particularly striking in the case of OH/IR stars (see \cite{2005A&A...435..563C}). In the case of Mz\,3, the similar shape of the visibility curves allows us to attribute it without ambiguity to amorphous silicate. Mz\,3 is a well-studied PN, and the a-priori information on its large scale geometry and dust content severely limits the range of spatial and physical parameters of the detected disk. The number and quality of the interferometric data were enough to conduct a direct interpretation via a radiative transfer model of a passive disk. 

\subsection{A-priori knowledge on the disk of Mz\,3}

The large-scale spatial information originates essentially in the extended bipolar lobes (Santander-Garcia et al. 2004; \cite{2004AJ....128.1694G}), whose kinematic age is limited to less than 1500yrs.
The estimated inclination of the different structures ranges from 68\deg\ to 78\deg, and \cite{2004AJ....128.1694G} mention a PA of 10$\pm$2\deg for the lobes. 
The mid-IR study from \cite{2005AJ....129..969S} also provides some important information: the mass of warm dust inside the core is rather limited (minimum of $1.1\times10^{-6}M_{\odot}$, optically thin environment and small carbon grain hypothesis), and the ratio between the mid-IR and visible luminosity suggests a modest opening angle ($\leq$24\deg).

\begin{table}
	\centering
		\begin{tabular}{ccc}
		\hline
		 & Parameters & Parameter range\\
		\hline
		$T_{eff}$ (K) & $35000$  & -\\
		Luminosity $(L_{\odot})$ & $10000$& -\\
		Distance (kpc)& $1.4$&  -\\
		\hline
		Inclination ($^{\circ}$) & $74$  & 3\\
		PA angle ($^{\circ}$)& 5  &5\\
		Inner radius (AU) &  9 & 1\\
		Outer radius (AU) &  500& No constraint \\
		$\alpha$ & 2.4  & 0.1\\
		$\beta$ & 1.02 & 0.02\\
		$h_{100AU}$ (AU)&17 &2\\
		Density (g.cm$^{-3}$)&  2.7 & - \\
		Dust mass $(M_{\odot})$&  $9\times 10^{-6}$ & $2\times 10^{-6}$\\
 		\hline
		 
		\end{tabular}
	\caption{Model parameters and those from the model fitting. }
	\label{tab:modparam}
\end{table}

\subsection{A passive disk model}
We made use of the 3D continuum radiative transfer code MC3D (Wolf 2003; see also Wolf et al. 1999). It is based on the Monte-Carlo method and solves the radiative transfer problem self-consistently. MC3D is designed to simulate dust temperatures in arbitrary dust/electron configurations and the resulting observables: spectral energy distributions, wavelength-dependent images, and polarization maps. We use a classical model of a dusty stratified disk (Shakura \& Sunyaev 1973; Wood et al. 2002). The dust density follows a 2D law (radial and vertical dimensions):
$$		\rho(r,z) = \rho_{0}\left(\frac{R_{\star}}{r}\right)^{\alpha} exp\left[-\frac{1}{2}\left(\frac{z}{h(r)}\right)^{2}\right]$$	
where $r$ is the radial distance in the midplane of the disk, $\alpha$ the density parameter in the midplane,
	$R_{\star}$ the stellar radius and the disk scale height $h(r)$ is given by $h(r)=h_{0}\left(\frac{r}{R_{\star}}\right)^{\beta}$, where $h_{0}$ is the scale height at a given radial distance from the star and $\beta$ is the vertical density parameter.\\

We assume the standard interstellar grain size distribution (Mathis et al. 1977): $\frac{dn(a)}{da}\  \alpha\  a^{-3.5}$ where $a$ is the dust grain radius. Grain radii extend from $0.05$ to $1 \mu$m and we consider the dust grains to be homogeneous spheres. The MIDI data are not suited to constrain the dust size distribution efficiently, so that the dust grain sizes are assumed and kept fixed in the fitting process. Similarly, the outer radius of the structure is not well-defined by these N band observations, so it is arbitrarily limited to 500\,AU (0.35$\arcsec$ at 1.4\,kpc).
The outputs from the code are a large aperture SED, 15 images (2mas per pixel) in the wavelength range 8-13.5~$\mu$m, and by applying 2D Gaussian apertures to the images (FWHM from 210mas to 350mas), we get the MIDI spectrum and the spectrally dispersed visibilities (taking the limited spectral resolution into account).
To generate the visibility signal, the images are first rotated by the PA of the object in the sky and then collapsed as 1D flux distributions in the direction of the baselines. The visibility for each wavelength is the value of the Fourier transform of these 1D vectors, normalized to the zero frequency value. 
	
The distance, luminosity, and temperature of the central star being uncertain by a large factor, we fixed them to plausible values close to the ones in \cite{2003MNRAS.342..383S}, trying to find an adequate fit of the SED and visibilities with them. It happened that such a good fit was found rapidly, and we did not try to explore the parameter space for the central source further, because we think that the current data are not sensitive enough to distinguish one model from another. With such a large inclination, the $\beta$ and $h$ parameters are well-fitted using the balance of visibilities between baselines perpendicular and parallel to the structure.

\section{Results and discussion}
The MIDI observations provide evidence of a flat, nearly edge-on disk primarily composed of amorphous silicate whose parameters are described in Table 2. The inner rim is at sublimation temperature and the disk is optically thick ($\tau_N\sim3.5$). The dust chemistry agrees with \cite{2005A&A...431..523B} and \cite{2005MNRAS.362.1199C} who found  evidence in the ISO spectra of weak signatures of amorphous silicate and no obvious signature of carbon dust in the form of PAHs, in accordance with the $C/O$ ratio of 0.83 determined by \cite{2002MNRAS.337..499Z} in the gas phase. The match of the SED and the visibility curves is impressively good showing that such a simple model of a passive disk can account for these observations. That the silicate is in the form of amorphous grains is important, since the long-lived disks found in binary post-AGBs exhibit strong signatures of {\it crystalline} silicate (\cite{2007A&A...467.1093D, 2007BaltA..16..145D}). 
This suggests that the disk may be relatively young, in line with the youth of the PN.
The dust mass stored in the disk, estimated to be $1\times 10^{-5} M_{\odot}$ is two orders of magnitude below the mass inferred in the lobes ($2.6\times 10^{-3} M_{\odot}$ from \cite{2005AJ....129..969S}). The low mass of the disk seems insufficient for pinching the waist of the bipolar nebula hydrodynamically, pointing instead to an essentially polar ejection process. The small disk opening angle and the limited angle sustained by the lobes also favor such a scenario. 
The detection of a bright X-ray core and a possible jet, reported by \cite{2003ApJ...591L..37K}, is probably a good indication that a compact accretion disk is hidden from direct view by the more extended dusty disk presented here.
Using near-IR color-color diagram and the similarity between the locus of symbiotic stars and Mz\,3, \cite{2001A&A...377L..18S} suggested that a Mira should be hidden in the center of the system. Our preferred interpretation is that this kind of diagram is useful for probing {\it dusty disks}, but is not convincing evidence of a Mira star. Only the temporal variability of the near-IR flux would represent such evidence but it is currently lacking  for Mz\,3. The constraint provided by our observations is that the orbit of the system must be well within the inner radius of the disk, at a distance of at most 1-2 AU from the primary, which is typical of S-type symbiotics (\cite{2003ASPC..303....9M}) but much too small for D-type symbiotic Miras. Therefore a Mira seems improbable, but a less luminous cool giant companion might remain hidden by the disk. The constraints from the SED are tight, limiting the cool companion luminosity to less than 100-150L$_\odot$ (\cite{2003MNRAS.342..383S}). 
This object could be a young link to the binary post-AGBs (\cite{2006A&A...448..641D}) that probably 
experienced non conservative mass loss. The current estimate of the mass stored in the Mz\,3 disk is in line with the mass of long-lived (crystalline) disks found in many of the binary post-AGBs.

\begin{acknowledgements}
A. Collioud is warmly thanked for his precious helping handling the MC3D code. 
\end{acknowledgements}

\end{document}